\documentclass[journal]{IEEEtran}
\usepackage{color}
\usepackage{float}
\usepackage{tabularx,colortbl}
\usepackage{setspace}
\usepackage{cite,graphicx,amsmath,psfrag,bm}
\usepackage{subfigure}
\usepackage[usenames,dvipsnames]{xcolor}
\usepackage{mathabx}
\usepackage{pifont}
\usepackage{cancel}
\usepackage{epstopdf}
\usepackage{amsmath}
\usepackage[thinlines]{easytable}
\usepackage{csquotes}
\usepackage{framed}
\usepackage{stmaryrd}
\usepackage{xr}
\usepackage{cite}

\usepackage[process=auto]{pstool}
\usepackage[normalem]{ulem}
\definecolor{red}{RGB}{237,28,36}
\definecolor{blue}{RGB}{57,83,164}
\definecolor{beige}{RGB}{250,164,26}
\definecolor{magenta}{RGB}{185,82,159}
\definecolor{green}{RGB}{0,161,75}
\definecolor{orange}{RGB}{241,90,41}

\newcommand{\figref}{Fig.~\ref}

\newcommand{\ve}[1]{\mathbf{#1}}
\newcommand{\ves}[1]{\boldsymbol{#1}}
\newcommand{\uve}[1]{\mathbf{\hat{{#1}}}}
\newcommand{\dya}[1]{\bar{\bar{#1}}}
\newcommand{\tx}[1]{\text{#1}}

\newcommand*{\TakeFourierOrnament}[1]{{%
\fontencoding{U}\fontfamily{futs}\selectfont\char#1}}
\newcommand*{\danger}{\TakeFourierOrnament{66}}

\makeatletter
\newcommand*{\addFileDependency}[1]{
  \typeout{(#1)}
  \@addtofilelist{#1}
  \IfFileExists{#1}{}{\typeout{No file #1.}}
}
\makeatother

\newcommand*{\myexternaldocument}[1]{%
    \externaldocument{#1}%
    \addFileDependency{#1.tex}%
    \addFileDependency{#1.aux}%
}

\myexternaldocument{AWPL_Nonreciprocity_Part_I}

\ifCLASSINFOpdf
\else
\fi

\begin{document}

\title{What is Nonreciprocity? -- Part II\vspace{-3mm}}

\author{Christophe~Caloz,~\IEEEmembership{Fellow,~IEEE},
Andrea Al\`{u},~\IEEEmembership{Fellow,~IEEE},
Sergei Tretyakov,~\IEEEmembership{Fellow,~IEEE},
Dimitrios Sounas,~\IEEEmembership{Senior Member,~IEEE},
Karim Achouri,~\IEEEmembership{Student Member,~IEEE},
and
Zo\'{e}-Lise Deck-L\'{e}ger\vspace{-12mm}
\thanks{
C.~Caloz and Z-L.~Deck-L\'{e}ger are with
Polytechnique Montr\'{e}al, Montr\'{e}al, QC H3T-1J4, Canada (e-mail: christophe.caloz@polymtl.ca).
A.~Al\`{u} is with the CUNY Advanced Science Research Center, NY 10031, USA. D.~Sounas and A.~Al\`{u} are with
the University of Texas at Austin, Austin, TX 78712, USA.
S.~Tretyakov is with
Aalto University, Aalto, FI-00076, Finland.
K.~Achouri is with
EPFL, Lausanne, CH-1015, Switzerland.}
}
\markboth{IEEE Antennas Wireless Propagation Letters,~Vol.~17, 2018}
{Shell \MakeLowercase{\textit{et al.}}: Bare Demo of IEEEtran.cls for IEEE Journals}

\maketitle

\begin{abstract}
This paper is the second part of a two-part paper on \emph{Electromagnetic (EM) Nonreciprocity (NR)}.
Part~I has defined NR, pointed out that linear NR is a stronger form of NR than nonlinear (NL) NR, explained EM Time-Reversal (TR) Symmetry (TRS) Breaking (TRS-B), described linear Time-Invariant (TI) NR media, generalized the Lorentz reciprocity theorem for NR, and provided a physical interpretation of the resulting Onsager-Casimir relations~\cite{Caloz_AWPL_NR_I_2018}. This part first explains the TR specificity of lossy and open systems. Next, it proposes an extended version of the S-parameters for \emph{all NR} systems. Then, it presents the fundamentals of linear-TI (LTI) NR, linear Time-Variant (LTV) Space-Time (ST) modulated NR and NL NR systems. Finally, it addresses confusions between with systems.
\end{abstract}
\vspace{-1mm}
\vspace{-3mm}

\IEEEpeerreviewmaketitle

\section{Introduction}
\vspace{-1mm}

Part~I~\cite{Caloz_AWPL_NR_I_2018} has introduced the overall paper and covered the topics enumerated in the abstract. This second part, after subdividing the linear category of NR systems (Tab.~\cite{Caloz_AWPL_NR_I_2018}.\ref{tab:classification}) into Linear Time-Invariant (LTI) and Linear Time-Variant (LTV)\footnote{Consider the isotropic dielectric medium system: $\ve{D}=\epsilon\ve{E}$. The system is clearly \emph{NL} if $\epsilon=\epsilon(\ve{E})$, since it excludes superposition because $\epsilon=\epsilon(\ve{E}_1+\ve{E}_2)$~\cite{Boyd_2008}. However, we call here \emph{linear}-TV a system with $\epsilon=\epsilon(t)\neq\epsilon(\ve{E})$ insofar such a system \emph{supports superposition}, although it is often called NL due to the generation of new frequencies.} systems, expands the concepts of Part~I to lossy and open systems, extends the scattering parameter concept from LTI systems to LTV Space-Time (ST) and NL systems, presents the fundamental NR characteristics and devices for these three types of systems, and finally warns against fallacious similarities between NR and other forms of asymmetric transmission.

\vspace{-2mm}
\begin{table}[h]
\footnotesize
\centering
\begin{tabular}{@{\hskip -1mm}c@{\hskip -0.2mm}|c@{\hskip -0.06mm}cc}
& \multicolumn{2}{c}{\textbf{LINEAR}} & \hspace{-2mm}\textbf{NONLINEAR}\hspace{-2mm} \\
\vspace{-0.3mm}& \textsc{TI} & \textsc{TV (Space-Time)} & \\
\vspace{-0.7mm}
 & \scriptsize{$\tilde{\dya{\chi}}(\ve{B}_0){\hskip -1mm}\neq{\hskip -1mm} \tilde{\dya{\chi}}^T{\hskip -1mm}(\ve{B}_0)$}\:\phantom{.}
 & \scriptsize{$\dya{\chi}(\ve{v}_0){\hskip -1mm}\neq{\hskip -1mm}\dya{\chi}^T{\hskip -1mm}(\ve{v}_0)$}
 & \hspace{-7mm}\scriptsize{$\dya{\chi}(\ve{E}){\hskip -1mm}\neq{\hskip -1mm}\dya{\chi}^T(\ve{E})$}\hspace{-7mm} \\
\\[-0.9em]
\hline
\\[-0.8em]
\begin{minipage}{0.135\textwidth}\centering
\textsc{Time Reversal} \\
\vspace{-0.8mm}Secs.~\cite{Caloz_AWPL_NR_I_2018}.\ref{sec:TRS}-\ref{sec:TRSB_example}
\end{minipage}
& \ding{52}& \ding{52} & \ding{52} \\
\\[-0.7em]
\begin{minipage}{0.135\textwidth}\centering
\textsc{Onsager-Casimir} \\
\vspace{-0.8mm}Eqs.~\cite{Caloz_AWPL_NR_I_2018}.\eqref{eq:Onsager_Casimir_rel}
\end{minipage}
& \ding{52} & \ding{52}  & \ding{52} \\
\\[-0.7em]
\begin{minipage}{0.135\textwidth}\centering
\textsc{Lorentz NR} \\
\vspace{-0.8mm}Secs.~\cite{Caloz_AWPL_NR_I_2018}.\ref{sec:Lin_NR_media}-\ref{sec:gen_rec_th}
\end{minipage}
& \ding{52} & \ding{56}  & \ding{56} \\
\\[-0.7em]
\begin{minipage}{0.135\textwidth}\centering
\textsc{Extended S-Par.} \\
\vspace{-0.8mm}Sec.~\ref{sec:scat_pat_mod}
\end{minipage}
& \ding{52} & \ding{52}\danger  & \ding{52}\danger\danger
\end{tabular}
\caption{Fundamental concept applicability to different NR systems.}
\label{tab:applicability}
\end{table}
\vspace{-4mm}

Table~\ref{tab:applicability}, with generic  susceptibility $\dya{\chi}$ ($\dya{\chi}^T$: opposite dir.), presents \textbf{\emph{Time Reversal Symmetry Breaking (TRS-B)} and \emph{Onsager-Casimir relations} as a common descriptor for all NR systems}, points out that the Lorentz theorem is applicable only to LTI systems, and indicates that the extended S-parameter apply to the three systems with growing restriction.

\section{Reciprocity despite TRS-B in Lossy Systems}\label{sec:Lossy_syst_paradox}
\vspace{-1mm}

\begin{figure}[h]
    \centering
        \psfragfig[width=0.92\linewidth]{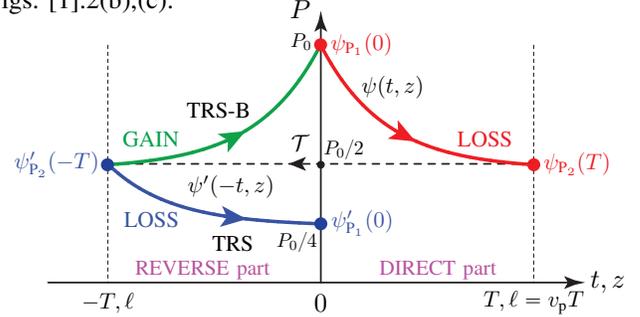}{
        \psfrag{t}[c][c]{$t,z$}
        \psfrag{P}[c][c]{$P$}
        \psfrag{0}[c][c]{$0$}
        \psfrag{T}[c][c][0.85]{$T,\ell=v_\tx{p}T$}
        \psfrag{M}[c][c][0.85]{$-T,\ell$}
        \psfrag{Q}[c][c][0.85]{$\psi(t,z)$}
        \psfrag{S}[c][c][0.85]{$\psi'(-t,z)$}
        \psfrag{Z}[c][c][0.75]{$P_0$}
        \psfrag{Y}[c][c][0.75]{$P_0/2$}
        \psfrag{X}[c][c][0.75]{$P_0/4$}
        \psfrag{N}[c][c][0.85]{\textcolor{red}{LOSS}}        \psfrag{L}[c][c][0.85]{\textcolor{blue}{LOSS}}
        \psfrag{G}[c][c][0.85]{\textcolor{green}{GAIN}}
        \psfrag{1}[c][c][0.85]{\textcolor{red}{$\psi_{\tx{P}_1}(0)$}}
        \psfrag{2}[c][c][0.85]{\textcolor{red}{$\psi_{\tx{P}_2}(T)$}}
        \psfrag{3}[c][c][0.85]{\textcolor{blue}{$\psi_{\tx{P}_2}'(-T)$}}
        \psfrag{4}[c][c][0.85]{\textcolor{blue}{$\psi_{\tx{P}_1}'(0)$}}
        \psfrag{B}[c][c][0.85]{TRS}
        \psfrag{C}[c][c][0.85]{TRS-B}
        \psfrag{D}[c][c][0.8]{\textcolor{magenta}{DIRECT part}}
        \psfrag{E}[c][c][0.8]{\textcolor{magenta}{REVERSE part}}
        \psfrag{R}[c][c][0.85]{${\cal T}$}}
        \vspace{-3mm}
        \caption{TRS-B in a lossy reciprocal waveguide of length $\ell$ (Sec.~\cite{Caloz_AWPL_NR_I_2018}.\ref{sec:TRS_breaking}). Assuming that the process under consideration is the propagation of a modulated pulse, we have $\ves{\Psi}'(-t)=[\psi_\tx{P$_1$}'(-t),\psi_\tx{P$_2$}'(-t)]^T
        \neq[\psi_\tx{P$_1$}(t),\psi_\tx{P$_2$}(t)]^T=\ves{\Psi}(-t)$, and in particular $\ves{\Psi}'(0)=[P_0/4,0]^T\neq[P_0,0]^T=\ves{\Psi}(0)$.}
        \vspace{-2mm}
   \label{fig:Loss_TRSB}
\end{figure}

Figure~\ref{fig:Loss_TRSB} monitors the process of EM wave propagation in a \textbf{\emph{lossy bias-less waveguide system}}. Let us see how such a system responds to TR (Secs.~\cite{Caloz_AWPL_NR_I_2018}.\ref{sec:TRS} to \cite{Caloz_AWPL_NR_I_2018}.\ref{sec:TRS_EM}) by applying the \textbf{TRS-B test} in Sec.~\cite{Caloz_AWPL_NR_I_2018}.\ref{sec:TRS_breaking}, as for the Faraday system in Figs.~\cite{Caloz_AWPL_NR_I_2018}.\ref{fig:Gyro_Rec_Med}(b),(c).

\vspace{-9mm}

In the direct part of the process, the wave is attenuated by dissipation as it propagates from port P$_1$ to port P$_2$ (red curve), say from $P_0$ to $P_0/2$ (3~dB loss). Upon TR, the propagation direction is reversed, \emph{and} loss is transformed into gain (Tab.~\cite{Caloz_AWPL_NR_I_2018}.\ref{tab:Field_Symmetries}). As a result, the wave propagates back from P$_2$ to P$_1$ \emph{and} its power level is restored (green curve), from $P_0/2$ to $P_0$. However, the \emph{system has been altered}. Maintaining lossy leads to further attenuation on the return trip, from $P_0/2$ to $P_0/4$ (6~dB loss), and hence \textbf{breaks TRS}. According to Sec.~\cite{Caloz_AWPL_NR_I_2018}.\ref{sec:TRS_breaking}, this would imply NR, which is at odds with the generalized reciprocity theorem (Sec.~\cite{Caloz_AWPL_NR_I_2018}.\ref{sec:gen_rec_th})!

This paradox originates in the looseness\footnote{This looseness has been tolerated because it is the \emph{ratio} definition of Sec.~\cite{Caloz_AWPL_NR_I_2018}.\ref{sec:NR_def_class}, and not its restricted \emph{level} form, that is commonly used in practice.} of the assumption that \textbf{``TRS/A is equivalent to reciprocity/NR''} in Sec.~\cite{Caloz_AWPL_NR_I_2018}.\ref{sec:TRS}. This assumption is, as pointed out in Fn.~\cite{Caloz_AWPL_NR_I_2018}.\ref{fn:loose_def}, stricto senso incorrect, as the equivalence \textbf{only holds in terms of absolute field levels} but not field ratios! In the case of loss, as just seen, the field ratios are equal [$(P_0/4)/(P_0/2)=0.5=(P_0/2)/P_0$], consistently with the general definition of reciprocity in Sec.~\cite{Caloz_AWPL_NR_I_2018}.\ref{sec:NR_def_class}, but the field levels are not ($P_0/4\neq P_0$), in contradiction with the definition of TR in Sec.~\cite{Caloz_AWPL_NR_I_2018}.\ref{sec:TRS}. In this sense, \textbf{\emph{a simple lossy system is perfectly reciprocal despite breaking TRS}}. This TRA may be seen as an expression of \textbf{\emph{thermodynamical macroscopic irreversibility}}\footnote{Consider for instance an empty metallic waveguide. The transfer of charges from the waveguide lossy walls results in EM energy being transformed into heat (Joule first law)~\cite{Casimir_1963}. In theory, a \emph{Maxwell demon}~\cite{Knott_1911}
could reverse the velocities of all the molecules of the system, which would surely reconvert that heat into EM energy. In this sense, \emph{all systems are microscopically reversible}, which is the fundamental assumption underpinning Onsager reciprocity relations~\cite{Onsager_1931_I,Onsager_1931_II,Casimir_1945,Landau_1997} (Fn.~\cite{Caloz_AWPL_NR_I_2018}.\ref{fn:Onsager}). However, such reconversion is prohibited by the second law of thermodynamics, which stipulates that \emph{the total entropy in an isolated system cannot decrease over time}. It would at the least require injecting energy from the outside of the system! So, such a lossy system is \emph{macroscopically -- and hence practically! -- irreversible}. Loss cannot be undone; it ever accumulates over time, as in \figref{fig:Loss_TRSB}.}.

\section{Open Systems and their TR-``Lossy'' Behavior}\label{sec:open_syst}
\vspace{-1mm}

\begin{figure}[h]
\centering
        \psfragfig[width=0.85\linewidth]{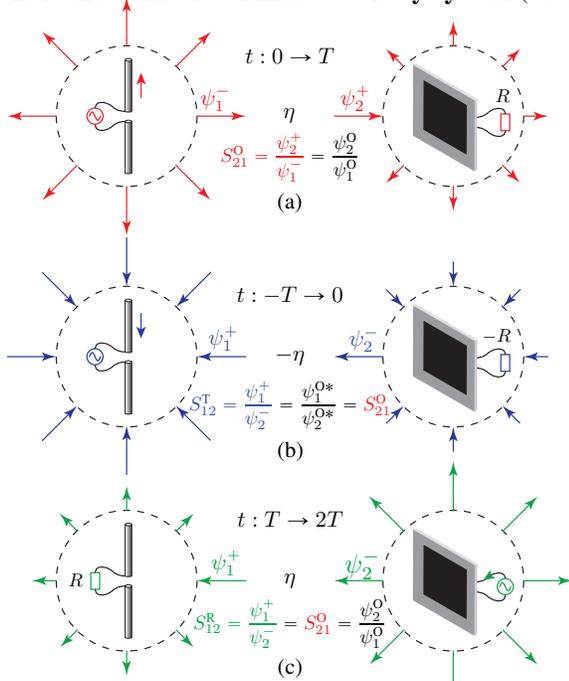}{
        \psfrag{a}[c][c][0.85]{(a)}
        \psfrag{b}[c][c][0.85]{(b)}
        \psfrag{c}[c][c][0.85]{(c)}
        \psfrag{1}[c][c][0.85]{$\eta$}
        \psfrag{2}[c][c][0.85]{$-\eta$}
        \psfrag{3}[c][c][0.85]{$\eta$}
        \psfrag{4}[c][c][0.7]{$R$}
        \psfrag{5}[c][c][0.7]{$-R$}
        \psfrag{6}[c][c][0.7]{$R$}
        \psfrag{h}[c][c][0.85]{$\textcolor{red}{\psi_1^-}$}
        \psfrag{i}[c][c][0.85]{$\textcolor{red}{\psi_2^+}$}
        \psfrag{j}[c][c][0.85]{$\textcolor{blue}{\psi_1^+}$}
        \psfrag{k}[c][c][0.85]{$\textcolor{blue}{\psi_2^-}$}
        \psfrag{l}[c][c][0.85]{$\textcolor{green}{\psi_1^+}$}
        \psfrag{m}[c][c]{$\textcolor{green}{\psi_2^-}$}
        \psfrag{e}[c][c][0.73]{$\textcolor{red}{S_{21}^\text{O}=\dfrac{\psi_2^+}{\psi_1^-}}=\dfrac{\psi_2^\text{O}}{\psi_1^\text{O}}$}
        \psfrag{f}[c][c][0.73]{$\textcolor{blue}{S_{12}^\text{T}=\dfrac{\psi_1^+}{\psi_2^-}}=\dfrac{\psi_1^{\text{O}*}}{\psi_2^{\text{O}*}}=\textcolor{red}{S_{21}^\text{O}}$}
        \psfrag{g}[c][c][0.73]{$\textcolor{green}{S_{12}^\text{R}=\dfrac{\psi_1^+}{\psi_2^-}}=\textcolor{red}{S_{21}^\text{O}}=\dfrac{\psi_2^\text{O}}{\psi_1^\text{O}}$}
        \psfrag{O}[c][c][0.85]{$t:0\rightarrow T$}
        \psfrag{T}[c][c][0.85]{$t:-T\rightarrow 0$}
        \psfrag{R}[c][c][0.85]{$t:T\rightarrow 2T$}
        }\vspace{-5mm}
        \caption{TRA and apparent (restricted) nonreciprocity of an open system composed of a dipole antenna and a patch antenna in free space.}
\label{fig:Diff_TRS_Rec_Open}
\vspace{-8mm}
\end{figure}

Consider the two-antenna open system in \figref{fig:Diff_TRS_Rec_Open}, showing the original [Fig.~\ref{fig:Diff_TRS_Rec_Open}(a)], TR [Fig.~\ref{fig:Diff_TRS_Rec_Open}(b)] and reciprocal [Fig.~\ref{fig:Diff_TRS_Rec_Open}(c)] problems. The nature of the system is clearly altered upon TR, where the intrinsic impedance of the surrounding medium becomes negative (Tab.~\cite{Caloz_AWPL_NR_I_2018}.\ref{tab:Field_Symmetries}). This results from the fact that \textbf{the \emph{radiated and scattered energy escaping the antennas} in the original problem is \emph{equivalent to loss} relatively to the \emph{two-port system}}. Such loss transforms into gain upon TR, as in Sec.~\ref{sec:Lossy_syst_paradox}, leading to fields emerging from infinity. Upon replacing TR by restricted-TR so as to avoid denaturing the system, one would find, as in the lossy case, reduced field levels but conserved field ratios\footnote{However, the restricted-TR problem is still \emph{distinct} from the reciprocity problem [Fig.~\ref{fig:Diff_TRS_Rec_Open}(c)] insofar as it does not reset the field level.}. \textbf{An open system is thus TR-wise similar to a lossy system} (Sec.~\ref{sec:Lossy_syst_paradox}).
\vspace{-6mm}

\section{Extended Scattering Parameter Modeling}\label{sec:scat_pat_mod}
\vspace{-1mm}

The lossy/open system paradox (Secs.~\ref{sec:Lossy_syst_paradox}/\ref{sec:open_syst}) and the common definition in Sec.~\cite{Caloz_AWPL_NR_I_2018}.\ref{sec:NR_def_class}, suggest describing \emph{LTI (non)reciprocal systems} in terms of \textbf{\emph{field ratios}}. This leads to the \textbf{\emph{scattering parameters} or \emph{S-parameters}}, introduced in quantum physics in 1937~\cite{Wheeler_1937}, used for over 70 years in microwave engineering~\cite{Dicke_1947,Marcuvitz_1951,Pozar_ME_2011}, extended to power parameters for arbitrary loads in the 1960ies~\cite{Kurokawa_1965,Pozar_ME_2011}, and to the cross-coupled matrix theory for topologically-coupled resonators in the 2000s~\cite{Atia_1972,Cameron_2007}. We shall attempt here an \textbf{\emph{extension} of these parameters to LTV and NL systems}.

\begin{figure}[h]
    \centering
        \psfragfig[width=0.7\linewidth]{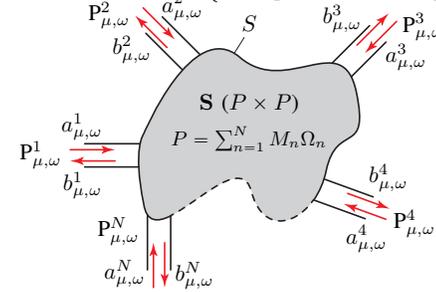}{
        \psfrag{S}[c][c][0.85]{\begin{minipage}{3cm}\centering $\ve{S}$ ($P\times P$) \\ \vspace{2mm} \small$P=\sum_{n=1}^N M_n\Omega_n$ \end{minipage}}
        \psfrag{a}[c][c][0.85]{$a^1_{\mu,\omega}$}
        \psfrag{b}[c][c][0.85]{$b^1_{\mu,\omega}$}
        \psfrag{c}[c][c][0.85]{$a^2_{\mu,\omega}$}
        \psfrag{d}[c][c][0.85]{$b^2_{\mu,\omega}$}
        \psfrag{e}[c][c][0.85]{$a^3_{\mu,\omega}$}
        \psfrag{f}[c][c][0.85]{$b^3_{\mu,\omega}$}
        \psfrag{g}[c][c][0.85]{$a^4_{\mu,\omega}$}
        \psfrag{h}[c][c][0.85]{$b^4_{\mu,\omega}$}
        \psfrag{i}[c][c][0.85]{$a^N_{\mu,\omega}$}
        \psfrag{j}[c][c][0.85]{$b^N_{\mu,\omega}$}
        \psfrag{O}[c][c][0.85]{$S$}
        \psfrag{1}[c][c][0.85]{P$^1_{\mu,\omega}$}
        \psfrag{2}[c][c][0.85]{P$^2_{\mu,\omega}$}
        \psfrag{3}[c][c][0.85]{P$^3_{\mu,\omega}$}
        \psfrag{4}[c][c][0.85]{P$^4_{\mu,\omega}$}
        \psfrag{N}[c][c][0.85]{P$^N_{\mu,\omega}$}
        }
        \vspace{-8.5mm}
        \caption{Arbitrary $P$-port network and \emph{extended} scattering matrix, $\ve{S}$.}
        \label{fig:Multiport}
        \vspace{-3mm}
\end{figure}

Figure~\ref{fig:Multiport} defines an \emph{extended arbitrary $P$-port network} as an electromagnetic structure delimited by a surface $S$ with $N$ \emph{waveguide terminals}, $T_n$, supporting each a number of mode-frequency\footnote{``Frequency'' here refers to the new frequency set (possibly infinite or continuous) in LTV and NL systems, practically restricted to discrete spectra.\label{fn:Sp_restr}} ports, P$_p=$P$^n_{\mu,\omega}$ with $p=1,2,\ldots,P$ (e.g. if $T_1$ is a waveguide with the $M_1=2$ modes TE$_{10}$ and TM$_{11}$ and the $\Omega_1=2$ frequencies $\omega$ and $2\omega$, it includes the $M_1\Omega_1=4$ ports P$_1=$P$^1_{\text{TE}_{10},\omega}$, P$_2=$P$^1_{\text{TE}_{10},2\omega}$, P$_3=$P$^1_{\text{TM}_{11},\omega}$ and P$_4=$P$^1_{\text{TM}_{11},2\omega}$.) The transverse fields in the waveguides have the \emph{frequency-domain form}~\cite{Marcuvitz_1951}, extended here to LTV and NL systems,\vspace{-1mm}
\begin{equation}\label{eq:WG_Modes}
\begin{Bmatrix}
\tilde{\ve{E}}_{t,p}(x,y,z) \\
\tilde{\ve{H}}_{t,p}(x,y,z)
\end{Bmatrix}\hspace{-0.5mm}
=\hspace{-0.5mm}
\left(a_p e^{-j\beta_p z}\pm b_p e^{+j\beta_n z}\right)
\hspace{-1mm}
\begin{Bmatrix}
\tilde{\ve{e}}_{t,p}(x,y) \\
\tilde{\ve{h}}_{t,p}(x,y)
\end{Bmatrix},
\vspace{-1.5mm}
\end{equation}
where $\oiint_S(\tilde{\ve{e}}_{t,p}\times\tilde{\ve{h}}_{t,q})\cdot\uve{n}ds=2\delta_{pq}$\footnote{This relation also applies when $p$ and $q$ differ only by frequency (same terminal/mode) assuming narrow-band, and hence independent, port detectors.}, and where $a_p$/$b_p$ ($p=1,\dots,P$) are the port input/output complex wave amplitudes, related by the \textbf{\emph{extended S-matrix}}, $\ve{S}$, as\vspace{-2mm}
\begin{equation}\label{eq:b_eq_Sa}
\ve{b}=\ve{S}\ve{a},
\quad\text{with}\quad
\begin{Bmatrix}
\ve{b}=[b_1,b_2,\ldots,b_P]^T \\
\ve{a}=[a_1,a_2,\ldots,a_P]^T
\end{Bmatrix}.
\vspace{-11mm}
\end{equation}

If the system is \textbf{\emph{linear}}, each entry of the matrix can be expressed by the simple transfer function $S_{ij}=b_i/a_j|_{a_k=0,k\neq j}$, which corresponds to the \textbf{conventional definition} of the $S$-parameters, except for the \textbf{frequency port definition extension in the ST (LTV) case}. If the system is \textbf{\emph{NL}}, then $S_{ij}=S_{ij}(a_1,a_2,\dots,a_P)$, and therefore \textbf{\emph{all} the (significant) input signals must be simultaneously present in the measurement} of the transfer function $S_{ij}$, as done in \emph{Broadband Poly-Harmonic Distortion (PHD)}, used in the Keysight microwave Nonlinear Vector Network Analyzer (NVNA)~\cite{Root_2005,Root_2008,PNAX_10_2017}\footnote{E.g. $S_{21}^\tx{junction}=b_2/a_1|_{a_2=a_3=0}\neq S_{21}(a_3)$; $S_{21}^\tx{mixer}=S_{21}^\tx{mixer}(a_3,\tx{LO})$.\label{fn:PNAX}}.

In a \emph{LTI medium}, the bianisotropic reciprocity relations~\cite{Kong_2008,Rothwell_2008}  excited by the two states $\ve{a}'$ and $\ve{a}''$ with responses $\ve{b}'$ and $\ve{b}''$ read now (sources are outside of the system.)~\cite{Supp_Mat_NR}\vspace{-1.2mm}
\begin{gather}
\nabla\hspace{-0.5mm}\cdot\hspace{-0.5mm}\left(\tilde{\ve{E}}'\hspace{-0.5mm}\times\hspace{-0.5mm}\tilde{\ve{H}}''
\hspace{-0.5mm}-\hspace{-0.5mm}\tilde{\ve{E}}''\hspace{-0.5mm}\times\hspace{-0.5mm}\tilde{\ve{H}}'\right)
\hspace{-0.5mm}=\hspace{-0.5mm}j\omega\left[
\tilde{\ve{E}}''\hspace{-0.5mm}\cdot\hspace{-0.5mm}\left(\tilde{\dya{\epsilon}}
\hspace{-0.5mm}-\hspace{-0.5mm}\tilde{\dya{\epsilon}}^T\right)\hspace{-0.5mm}\cdot\hspace{-0.5mm}\tilde{\ve{E}}'\right.
\hspace{-0.5mm}-\hspace{-0.5mm}\tilde{\ve{H}}''\hspace{-0.5mm}\cdot\hspace{-0.5mm}\left(\tilde{\dya{\mu}}\hspace{-0.5mm}\right.
\nonumber \\
\vspace{-8mm}
\left.-\hspace{-0.3mm}\tilde{\dya{\mu}}^T\right)\hspace{-0.5mm}\cdot\hspace{-0.5mm}\tilde{\ve{H}}'
\hspace{-0.5mm}+\hspace{-0.5mm}\tilde{\ve{E}}''\hspace{-0.5mm}\cdot\hspace{-0.5mm}\left(\tilde{\dya{\xi}}
\hspace{-0.5mm}+\hspace{-0.5mm}\tilde{\dya{\zeta}}^T\right)\hspace{-0.5mm}\cdot\hspace{-0.5mm}\tilde{\ve{H}}'\left.\hspace{-0.5mm}-\tilde{\ve{H}}''\hspace{-0.5mm}\hspace{-0.5mm}\cdot\hspace{-0.5mm}\left(\tilde{\dya{\zeta}}
\hspace{-0.5mm}+\hspace{-0.5mm}\tilde{\dya{\xi}}^T\right)\hspace{-0.5mm}\cdot\hspace{-0.5mm}\tilde{\ve{E}}'\right]\hspace{-0.5mm}=\hspace{-0.5mm}0.
\label{eq:gen_rec_the_for_S}
\end{gather}
Inserting the sum ($\sum_{p=1}^P$) of fields~\eqref{eq:WG_Modes} into this equation, taking the volume integral of the resulting relation, applying the Gauss theorem and the orthogonality relation, and using the Onsager-Casimir relations [Eqs.~\cite{Caloz_AWPL_NR_I_2018}.\eqref{eq:Onsager_Casimir_rel}],
yields~\cite{Supp_Mat_NR}\vspace{-1mm}
\begin{equation}
\vspace{-1.5mm}
\textstyle{\sum_p}(b_p'a_p''\hspace{-0.5mm}-\hspace{-1mm}a_p'b_p'')
\hspace{-0.5mm}=\hspace{-0.5mm}\ve{b}\ve{a}^{\prime\prime T}\hspace{-0.5mm}-\hspace{-0.5mm}\ve{a}\ve{b}^{\prime\prime T}
\hspace{-0.5mm}=\hspace{-0.5mm}\ve{a}'\ve{a}^{\prime\prime T}\left(\ve{S}^T\hspace{-0.5mm}-\hspace{-0.5mm}\ve{S}\right),
\vspace{-0.5mm}
\end{equation}
where~\eqref{eq:b_eq_Sa} has been used to eliminate $\ve{b}^{(\prime,\prime\prime)}$ in the last equality. This leads to the reciprocity condition $\ve{S}=\ve{S}^T$, and hence to the convenient \textbf{scattering-parameter NR condition}
\begin{equation}\label{eq:Sp_NR}
  \ve{S}\neq\ve{S}^T
  \;\text{or}\;\; \exists~(i,j)~|~
  S_{ij}\neq S_{ji},\,i=,1,2,\ldots,N,
\end{equation}
(e.g. $S_{21}\neq S_{12}$) which also applies to LTV and NL systems, although the current demonstration is restricted to LTI media.


In the microwave regime, these S-parameters can be directly measured (magnitude and phase) with a VNA~\cite{Pozar_ME_2011} or with an NVNA. In contrast, in the optical regime no specific instrumentation is available for that and a special setup, with nontrivial phase handling, is therefore required~\cite{Jalas_2013}.

\section{Energy Conservation - Thermodynamic Paradox}\label{sec:therm_resol_parad}
\vspace{-1.9mm}

\textbf{In a \emph{lossless system}, \emph{energy conservation} requires that the total output power equals the total input power}, or $\sum_{p=1}^P|b_p|^2=\sum_{p=1}^P|a_p|^2$ in \figref{fig:Multiport}, since no power is dissipated in the system. In terms of \textbf{S-matrix}, this requirement translates into the \textbf{unitary} relation $\ve{S}\ve{S}^\dagger=I$ ($I$: unit matrix). Many fundamental useful facts on multi-port systems straightforwardly follow from energy conservation (e.g.~\cite{Pozar_ME_2011}).

Some immediate \textbf{consequences for NR} are: 1)~\emph{A lossless \textbf{1-port system} can be only totally reflective}, from $|S_{11}|^2=1$, even it includes NR materials, contrary to claims in~\cite{Tsakmakidis_2017}. 2)~\emph{A \textbf{2-port system} can be magnitude-wise NR only if it is lossy}; specifically, a purely reflective 2-port isolator $\ve{S}=[0,0;1,1]$ is impossible, since energy conservation requires $b_2^2=a_1^2+a_2^2$, whereas the device would exhibit $b_2^2=a_1^2+a_2^2+2a_1a_2$, with the additional term $2a_1a_2$ that may bring the total energy to be larger than the input power, depending on the relative phases of $a_1$ and $a_2$. 3)~\emph{A lossless \textbf{2-port system} can still be NR in phase}, since $\ve{S}\ve{S}^\dagger=I$ does \emph{not} demand $\angle S_{21}=\angle S_{12}$~\cite{Zhang_TMTT_09_2015}. 4)~\emph{A lossless \textbf{3-port system} can be matched simultaneously at all port only if it is NR}, according to the system $\ve{S}\ve{S}^\dagger=I$~\cite{Pozar_ME_2011}.

The case 2) has raised much perplexity in the past\footnote{This started in 1885 with the comment by Rayleigh that the system composed of two Nicols sandwiching a magnetized dielectric would be ``inconsistent with the second law of thermodynamics''~\cite{Strutt_1885}, to be overruled by Rayleigh himself 16 years later~\cite{Rayleigh_1901}, in reaction to studies of Wiener~\cite{Wiener_1900}.}, and led to the so-called \textbf{\emph{``thermodynamic paradox''}}, according to which such a system would ever increase the temperature of the load at the passing end at the detriment of the load at the isolated end, hence violating the second law of thermodynamics, which prescribes heat transfer from hot to cold bodies. The paradox resurfaced in 1955, as Lax and Button pointed out the existence of lossless unidirectional eigenmodes in some ferrite-loaded waveguide structures~\cite{Lax_1955}, but it was resolved by Ishimaru, who showed that such a waveguide would necessarily support loss in its terminations, even in the limit of negligible material loss~\cite{Ishimaru_1962,Ishimaru_1990}.\vspace{-3mm}

\section{Linear-TI (LTI) NR Systems}\label{sec:lin_NR_comp}
\vspace{-1.9mm}

\emph{LTI NR systems} have the following characteristics: 1)~TRS-B by TR-odd \textbf{external bias} $\ve{F}_0$, which is often a \textbf{magnetic field, $\ve{B}_0$} [Fig.~\cite{Caloz_AWPL_NR_I_2018}.\ref{fig:Gyro_Rec_Med}(c)]; 2)~applicability, by linearity, to \textbf{arbitrary excitations and intensities -- \emph{strong NR}} (Tab.~\cite{Caloz_AWPL_NR_I_2018}.\ref{tab:classification}); 3)~\textbf{frequency conservation}, and hence \emph{unrestricted} frequency-domain description (Secs.~\cite{Caloz_AWPL_NR_I_2018}.\ref{sec:Lin_NR_media} and \cite{Caloz_AWPL_NR_I_2018}.\ref{sec:ser_rec_from_TRS}), applicability of the \textbf{Lorentz reciprocity theorem} (Sec.~\cite{Caloz_AWPL_NR_I_2018}.\ref{sec:gen_rec_th}) and of \textbf{S-parameters} (Sec.~\ref{sec:scat_pat_mod}); 4)~generally \textbf{based on LTI materials}~\cite{Lax_1962,Auracher_1975,Ishak_1988,Rodrigue_1988,Zvezdin_Kotov_1997,Adam_2002,Kodera_2009,Kodera_2010,Parsa_TAP_03_2011}, including 2DEGs and graphene~\cite{Sounas_APL_01_2011,Sounas_TMTT_04_2012,Chamanara_MWCL_07_2012,Chamanara_OE_05_2013,Sounas_APL_05_2013,Tamagnone_2014,Tamagnone_2016}, \textbf{or metamaterials}~\cite{Carignan_2009,Boucher_2009,Kodera_APL_07_2011,Carignan_2011,Kodera_AWPL_01_2012,Kodera_AWPL_12_2012,Wang_PNAS_2012,Sounas_TAP_01_2013,Kodera_TMTT_03_2013,Taravati_TAP_07_2017,Kodera_2018} (Sec.~\cite{Caloz_AWPL_NR_I_2018}.\ref{sec:Lin_NR_media}).

The main LTI NR systems are isolators, NR phase shifters and circulators~\cite{Lax_1962,Rodrigue_1988,Pozar_ME_2011}. \textbf{\emph{Isolators}} ($\ve{S}=[0,0;1,0]$)\footnote{They may be of \emph{resonance, field-displacement or matched-port-circulator} type, and may involve resistive sheets, quarter-wave plates or polarizing grids.} are typically used to shield equipment (e.g. VNA or laser) from detuning, interfering and even destructive reflections. \textbf{\emph{NR phase shifters}} ($\ve{S}=[0,e^{j\Delta\varphi};1,0]$)\footnote{They may be of \emph{latching (hard magnetic hysteresis) or Faraday rotation} (Sec.~\cite{Caloz_AWPL_NR_I_2018}.\ref{sec:TRSB_example}) type, and may involve quarter-wave plates.}, such as \emph{gyrators} ($\Delta\varphi=\pi$)~\cite{Tellegen_1948}, combine with couplers to form isolators or circulators, provide compact simulated inductors and filter inverters, and enable NR pattern and scanning arrays. \textbf{\emph{Circulators}} ($\ve{S}=[0,0,1;1,0,0;0,1,0]$)\footnote{They may be of \emph{4-port Faraday rotation or 3-port junction rotation} type.} are used for isolation, duplexing (radar and communication), and reflection amplifiers.\vspace{-2mm}

\section{Linear-TV (LTV) Space-Time (ST) NR Systems}\label{sec:ST_var_syst}
\vspace{-2mm}

\emph{LTV ST}\footnote{As will be seen, NR requires also spatial inversion symmetry breaking.} \emph{NR systems} have the following characteristics: 1)~TRS-B by TR-odd \textbf{external bias} ($\ve{F}_0$) \textbf{velocity, $\ve{v}_0$}; 2)~\textbf{strong NR}, as in Sec.~\ref{sec:lin_NR_comp}; 3)~\textbf{generation of new, possibly anharmonic, frequencies}, and hence \emph{restricted} applicability of \textbf{S-parameters} (Fn.~\ref{fn:Sp_restr}); 4)~\textbf{\emph{moving medium}} (moving matter, e.g. opto-mechanical)~\cite{Sommerfeld_1952,Pauli_1958,Kong_2008,Jackson_1998,Kippenberg_2007} \textbf{or \emph{modulated medium}} (moving perturbation, e.g. electro/acousto/NL-optic)\footnote{Moving and modulated media both produce Doppler shifts~\cite{Eden_Doppler_1992} and NR. In contrast, only the former supports Fizeau drag~\cite{Jackson_1998,Fizeau_1851} and bianisotropy transformation~\cite{Kong_2008}, and only the latter allows superluminality~\cite{Deck_PRB_03_2018}.}~\cite{Cassedy_1963,Saleh_Teich_FP_2007}; 5)~pulse/periodic and abrupt/smooth mat./pert. motion.

\begin{figure}[h]
\centering
\psfragfig*[width=0.78\linewidth]{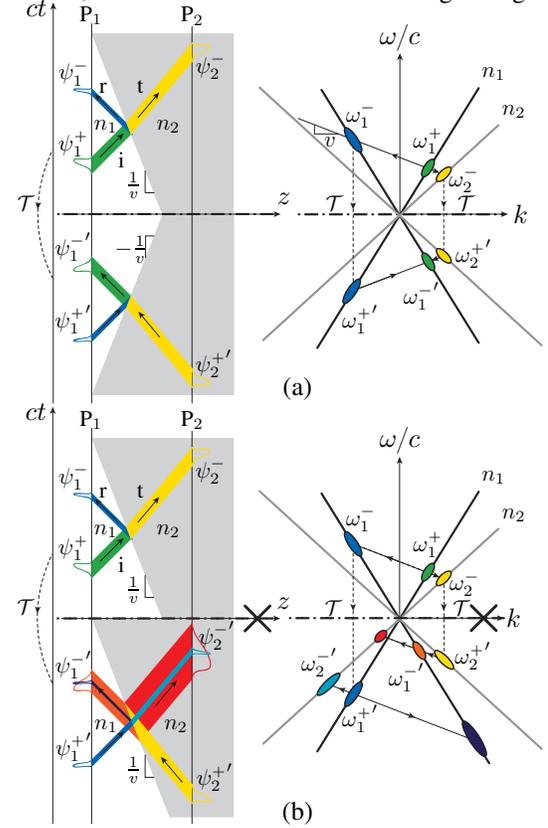}{
\psfrag{1}[c][c][0.85]{$n_1$}
\psfrag{2}[c][c][0.85]{$n_2$}
\psfrag{L}[c][c][0.85]{P$_1$}
\psfrag{R}[c][c][0.85]{P$_2$}
\psfrag{z}[c][c]{$z$}
\psfrag{Q}[c][c]{$ct$}
\psfrag{Y}[c][c]{(a)}
\psfrag{Z}[c][c]{(b)}
\psfrag{M}[c][c][0.85]{$n_1$}
\psfrag{N}[c][c][0.85]{$n_2$}
\psfrag{w}[c][c]{$\omega/c$}
\psfrag{k}[c][c]{$k$}
\psfrag{V}[c][c][0.85]{$\frac{1}{v}$}
\psfrag{v}[c][c][0.85]{$-\frac{1}{v}$}
\psfrag{s}[c][c][0.85]{$v$}
\psfrag{A}[c][c][0.85]{$\psi_1^+$}
\psfrag{B}[c][c][0.85]{$\psi_1^-$}
\psfrag{C}[c][c][0.85]{$\psi_2^-$}
\psfrag{a}[c][c][0.85]{${\psi_1^-}'$}
\psfrag{b}[c][c][0.85]{${\psi_1^+}'$}
\psfrag{c}[c][c][0.85]{${\psi_2^+}'$}
\psfrag{d}[c][c][0.85]{${\psi_2^-}'$}
\psfrag{E}[c][c][0.85]{$\omega_1^+$}
\psfrag{F}[c][c][0.85]{$\omega_1^-$}
\psfrag{G}[c][c][0.85]{$\omega_2^-$}
\psfrag{e}[c][c][0.85]{${\omega_1^-}'$}
\psfrag{f}[c][c][0.85]{${\omega_1^+}'$}
\psfrag{g}[c][c][0.85]{${\omega_2^+}'$}
\psfrag{h}[c][c][0.85]{${\omega_2^-}'$}
\psfrag{T}[c][c][0.8]{${\cal T}$}
\psfrag{i}[c][c][0.85]{i}
\psfrag{r}[c][c][0.85]{r}
\psfrag{t}[c][c][0.85]{t}
        }
\vspace{-3.5mm}
\caption{Step STM system, $t>0$. (a)~$t<0$: TRS. (b)~$t<0$: \mbox{TRA-NR}.}
\label{fig:Minkowski_TR}
\vspace{-2.5mm}
\end{figure}

As an illustration, Fig.~\ref{fig:Minkowski_TR} graphically depicts, using an extended Minkowski diagram~\cite{Deck_PRB_03_2018}, a \textbf{step ST-modulated (STM) system} with interface between media of refractive indices $n_1$ and $n_2$ ($n_2>n_1$) moving in the $-z$-direction with the constant velocity $\ve{v}_0=v\uve{z}$ ($v<0$) and excited by an incident (i) wave propagating in the $+z$-direction. Upon full-TR, $\ve{v}_0$ is reversed, which leads to identical \emph{Doppler shifts}~\cite{Eden_Doppler_1992} in the reflected (r) and transmitted (t) waves\footnote{Transitions follow frequency conservation lines in the moving frame~\cite{Deck_PRB_03_2018}.}, as in Fig.~\ref{fig:Minkowski_TR}(a), but alters the system. The unaltered system is TRA, and hence breaks TRS, which leads to the NR scattering in Fig.~\ref{fig:Minkowski_TR}(b)\footnote{Purely temporal modulation (horizontal interface)~\cite{Kalluri_ETVCM_2010,Fang_2012,Fang_2013} would clearly be insufficient for NR; spatial inversion symmetry breaking, provided here by the moving modulation, is also required to break TRS: $\dya{\chi}(\ve{v}_0)\neq\dya{\chi}^T(\ve{v}_0)$.}.\vspace{-5mm}

A great diversity of useful STM NR systems have been reported in recent years~\cite{Winn_1999,Dong_2008,Yu_2009,Yu_2009_2,Manipatruni_2009,Kang_2011,Yu_2011,Kamal_2011,Doerr_2011,Sounas_NATCOM_09_2013,Galland_2013,Qin_TMTT_2014,Sounas_ACS_2014,Estep_2014,Hadad_2015,Hadad_2016,Reiskarimian_2016,Taravati_TAP_02_2017,Taravati_PRB_03_2018}. They are all based on the production of \textbf{\emph{different traveling phase gradient in different directions}} and are, in that sense, more or less lumped/distributed~\cite{Pozar_ME_2011} \emph{variations of parametric systems developed by microwave engineers in the 1950ies}\footnote{The main difference is that the STM or \emph{parametric} systems of that time were developed mostly for amplifiers or mixers, rather than NR devices.}~\cite{Pierce_1950,Cullen_1958,Tien_1958,Tien_1958_2,Landauer_1960,LePage_1953,Franks_1960}.

Figure~\ref{fig:Minkowski_TR} shows a \textbf{\emph{NR metasurface reflector}} application~\cite{Shaltout_OME_11_2015,Hadad_2015} based on the ST modulation $n(x)=n_0+n_\tx{m}\cos(\beta_\tx{m}x+\omega_\tx{m}t)$, where $(\omega_\tx{m}/\beta_\tx{m})\uve{z}=\ve{v}_0$. The STM metasurface breaks reciprocity and hence provides a quite unique NR device by adding the spatial and temporal momenta $\ve{K}_\tx{MS}$ and $\omega_\tx{MS}$ to those of the incident wave\footnote{This system actually includes infinitely many ports, $S_{n,n+1}=0$ and $S_{n+1,n}$ with $\omega_{n+1}>\omega_n$ ($n=1,\ldots\infty$), but its \emph{functional} reduction in the figure is meaningful the power transfer beyond P$_3$ is of no interest.}.
\vspace{-5mm}
\begin{figure}[h]
    \centering
        \psfragfig*[width=0.95\linewidth]{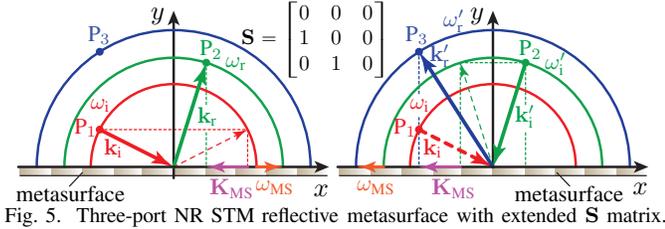}{
        \psfrag{x}[c][c]{$x$}
        \psfrag{y}[c][c]{$y$}
        \psfrag{1}[c][c][0.85]{\color{red}{$\ve{k}_\text{i}$}}
        \psfrag{2}[c][c][0.85]{\color{green}{$\ve{k}_\text{r}$}}
        \psfrag{3}[c][c][0.85]{\color{green}{$\ve{k}_\text{i}'$}}
        \psfrag{4}[c][c][0.85]{\color{blue}{$\ve{k}_\text{r}'$}}
        \psfrag{5}[c][c][0.85]{\color{magenta}{$\ve{K}_\text{MS}$}}
        \psfrag{6}[c][c][0.85]{\color{red}{$\omega_\text{i}$}}
        \psfrag{7}[c][c][0.85]{\color{green}{$\omega_\text{r}$}}
        \psfrag{8}[c][c][0.85]{\color{green}{$\omega_\text{i}'$}}
        \psfrag{9}[c][c][0.85]{\color{blue}{$\omega_\text{r}'$}}
        \psfrag{d}[c][c][0.85]{\color{orange}{$\omega_\text{MS}$}}
        \psfrag{h}[c][c][0.85]{\color{orange}{$\omega_\text{MS}$}}
        \psfrag{a}[c][c]{}
        \psfrag{b}[c][c]{}
        \psfrag{k}[c][c][0.85]{\color{red}{P$_1$}}
        \psfrag{l}[c][c][0.85]{\color{green}{P$_2$}}
        \psfrag{m}[c][c][0.85]{\color{blue}{P$_3$}}
        \psfrag{M}[c][c][0.85]{metasurface}
        \psfrag{S}[c][c][0.8]{$\ve{S}=\begin{bmatrix}0&0&0\\1&0&0\\0&1&0\end{bmatrix}$}
        }
        \vspace{-8mm}
        \caption{Three-port NR STM reflective metasurface with extended $\ve{S}$ matrix.}
        \label{fig:ST_Metasurface}
        \vspace{-8mm}
\end{figure}

\section{Nonlinear (NL) NR Systems}\label{sec:nonlin_syst}
\vspace{-2mm}

\emph{NL NR} systems have the following characteristics: 1)~TRS-B by \textbf{\emph{spatial asymmetry} and \emph{NL self-biasing}} (NL triggering by wave)~\cite{Naguleswaran_1998,Trzeciecki_2000}; 2)~limitation to \textbf{restricted excitations, intensities and isolation -- \emph{weak NR}} (Tab.~\cite{Caloz_AWPL_NR_I_2018}.\ref{tab:classification}); 3)~\textbf{generation of new, only harmonic, frequencies, and inapplicability of superposition}, and hence \emph{very restricted} applicability of \textbf{S-parameters} (Fns.~\ref{fn:Sp_restr} and~\ref{fn:PNAX}); 4)~diversity of TRS-B approaches.

\vspace{-6mm}
\begin{figure}[h]
    \centering
        \psfragfig[width=0.92\linewidth]{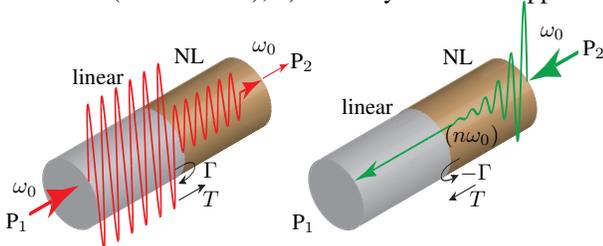}{
        \psfrag{N}[l][l][0.85]{NL}
        \psfrag{L}[l][l][0.85]{linear}
        \psfrag{X}[c][c][0.85]{P$_1$}
        \psfrag{Y}[c][c][0.85]{P$_2$}
        \psfrag{w}[c][c][0.85]{$\omega_0$}
        \psfrag{z}[c][c][0.85]{$(n\omega_0)$}
        \psfrag{R}[c][c][0.85]{$\Gamma$}
        \psfrag{r}[c][c][0.85]{$-\Gamma$}
        \psfrag{T}[c][c][0.85]{$T$}
        }
        \vspace{-5mm}
        \caption{Principle of nonlinear nonreciprocity with the asymmetric cascade of a linear medium, $\epsilon_\tx{L}=\epsilon_1$, and a nonlinear lossy medium, $\epsilon_\tx{NL}=\epsilon_2+[\epsilon'(E)-j\epsilon''(E)]$, with high mismatch, $|\Gamma|=|(\sqrt{\epsilon_\tx{NL}}-\sqrt{\epsilon_\tx{L}})/(\sqrt{\epsilon_\tx{NL}}+\sqrt{\epsilon_\tx{L}})|\gg 0$.}
   \label{fig:NL_NR}
   \vspace{-3mm}
\end{figure}

Figure~\ref{fig:NL_NR} shows a simple way to achieve NL NR \textbf{by pairing a linear medium and a NL lossy medium}. The two media are strongly \emph{mismatched}, with reflection~$\Gamma$. A wave injected at port P$_1$  experiences a transmittance of $|T|^2\hspace{-1mm}=\hspace{-1mm}1\hspace{-1mm}-\hspace{-1mm}|\Gamma|^2\hspace{-1mm}\ll\hspace{-1mm}1$, yielding a much smaller field level in the NL medium. If this level is insufficient to trigger NL loss, all the power transmitted through the interface ($|T|^2$) reaches port P$_2$, so that $|S_{21}|\hspace{-1mm}=\hspace{-1mm}|T|$ and $|S_{11}|\hspace{-1mm}=\hspace{-1mm}|\Gamma|$. The same wave injected at P$_2$, assuming sufficient intensity to trigger NL loss, undergoes exponential attenuation $e^{-\alpha\ell_\tx{NL}}$ ($\ell_\tx{NL}$: NL length), so that $|S_{12}|\hspace{-1mm}=\hspace{-1mm}|T|e^{-\alpha\ell_\tx{NL}}\hspace{-1mm}\approx\hspace{-1mm}0$. The system is thus NR, but it is a \textbf{\emph{pseudo}-isolator}\footnote{The device is also \emph{not a diode}~\cite{Sze_Ng_PSD_2006}, whose NR consists in different forward/backward spectra due to positive/negative wave cycle clipping.}: 1)~it is \textbf{restricted to a small range of intensities}; 2)~it works \textbf{only for one excitation direction} ($\tx{P}_1\hspace{-1mm}\rightarrow\hspace{-1mm}\tx{P}_2$ \emph{or} $\tx{P}_2\hspace{-1mm}\rightarrow\hspace{-1mm}\tx{P}_1$) \textbf{at a time}, since the $\tx{P}_2\hspace{-1mm}\rightarrow\hspace{-1mm}\tx{P}_1$ wave would trigger NL loss and hence also extinct the $\tx{P}_1\hspace{-1mm}\rightarrow\hspace{-1mm}\tx{P}_2$ wave\footnote{This precludes most of the applications of real isolators (Sec.~\ref{sec:lin_NR_comp})~\cite{Shi_2015}.}; 3)~it often suffers from \textbf{poor isolation} ($|S_{21}|\hspace{-0.3mm}/\hspace{-0.3mm}|S_{12}|\hspace{-0.5mm}=\hspace{-0.5mm}e^{\alpha\ell_\tx{NL}}$) and \textbf{poor isolation to insertion loss ratio} ($(|S_{21}|\hspace{-0.3mm}/\hspace{-0.3mm}|S_{12}/|)/|S_{11}|\hspace{-0.5mm}=\hspace{-0.5mm}e^{\alpha\ell_\tx{NL}}/|\Gamma|$)\footnote{These two parameters are typically smaller than $20$dB/$25$dB in NR NL structures, whereas they commonly exceed $45$dB/$50$dB in NR LTI isolators.}; 4)~it is \textbf{reciprocal to noise}~\cite{Shi_2015}.

Ingenious variations of the NL NR device in \figref{fig:NL_NR} have been reported~\cite{Gallo_1999,Kivshar_2010,Lepri_2011,Fan_2011,Fan_2012,Bender_2013,Wang_2013,Chang_2014,Nazari_2014,Engheta_2015,Sounas_2018,Sounas_AWPL_2018}. Some of them mitigate some of the aforementioned issues, but these improvements are severely restricted by fundamental limitations of NL NR~\cite{Jalas_2013,Shi_2015}.
\vspace{-4mm}

\section{Distinction with Asymmetric Propagation}\label{sec:dist_asym}
\vspace{-2mm}

\begin{figure}[h]
    \centering
        \psfragfig[width=0.7\linewidth]{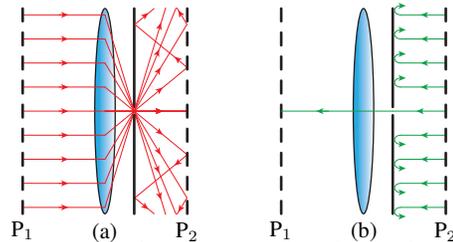}{
        \psfrag{a}[c][c][0.85]{(a)}
        \psfrag{b}[c][c][0.85]{(b)}
        \psfrag{x}[c][c][0.85]{P$_1$}
        \psfrag{y}[c][c][0.85]{P$_2$}
        }
        \vspace{-8mm}
        \caption{Asymmetric \emph{reciprocal} system formed by a lens and a mirror sandwiched between two antenna arrays under (a)~left and (b)~right excitations.}
   \label{fig:Lens_Hole_Asymmetry}
   \vspace{-7mm}
\end{figure}

\textbf{A NR system is a system that exhibits TRA field ratios between well-defined ports} [Eq.~\eqref{eq:Sp_NR}], which is possible only under \textbf{external biasing (linear NR) \emph{or} self-biasing plus spatial asymmetry (NL NR)}. Any system not satisfying this condition is \emph{necessarily reciprocal}, despite possible \textbf{\emph{fallacious asymmetries in transmission}}\cite{Fang_1996,Wang_OE_2011,Feng_Science_2011,Fan_Science_2012,Wang_SR_2012,Jalas_2013,Maznev_2013,Shi_2015,Fernandez_AWPL_2018}. For instance, the system in \figref{fig:Lens_Hole_Asymmetry} exhibits \textbf{asymmetric ray propagation}, but it is fully reciprocal since only the horizontal ray gets transmitted between the array ports, the $\tx{P}_1\rightarrow\tx{P}_2$ oblique rays symmetrically canceling out on the right array due to opposite phase gradients. Other deceptively NR cases include: \textbf{asymmetric field rotation-filtering} (e.g. $\pi/2$ rotator $+$ polarizer), where $S_{21}^{yx}\neq S_{21}^{xy}$, but $S_{21}^{yx}=S_{12}^{xy}$; \textbf{asymmetric waveguide junction} (e.g. step width variation), with full transmission to larger side distributed over multi-modes and small transmission from the same mode to the smaller (single-mode) side, but reciprocal mode-to-mode transmission (e.g. $S_{51}=S_{15}$, 1: small side single-mode port, 2: large side mode/port 5)~\cite{Jalas_2013}; \textbf{asymmetric mode conversion} (e.g. waveguide with nonuniform load), where even mode transmits in opposite directions with and without excitation of odd mode, without breaking reciprocity, since $S_{21}^\tx{ee}=S_{12}^\tx{ee}$ and $S_{21}^\tx{oe}=0=S_{12}^\tx{eo}$~\cite{Fan_Science_2012}. In all cases, reciprocity is verified upon exchanging the source and detector.
\vspace{-5mm}

\section{Conclusion}
\vspace{-2mm}

This paper has presented, in the context of novel magnetless NR systems aiming at repelling the frontiers of NR technology, a global perspective of NR, with the following conclusions: 1)~NR systems may be classified into linear (TI and TV-ST) and NL systems, based on TRS-B by external biasing and self-biasing plus spatial asymmetry, respectively; 2)~NL NR is a weaker form of NR than linear NR, as it suffers from restricted intensities, one-way-at-a-time excitations, and poor isolation and high insertion loss; 3)~lossy and open systems, although reciprocal in terms of field ratios, are TRS-A, due macroscopic irreversibility; 4)~S-parameters are advantageously generalized to all types of NR systems, with some restrictions; 5)~Care must be exercised to avoid confusing asymmetric transmission with NR in some fallacious systems. Nonreciprocity is a rich and fascinating concept, that has already and will continue to open new scientific and technological horizons.

\vspace{-1.5mm}

\clearpage

\bibliographystyle{IEEEtran}
\bibliography{IEEEabrv,AWPL_Nonreciprocity}

\end{document}